\documentclass[fleqn,usenatbib]{mnras}

\usepackage{newtxtext,newtxmath}

\usepackage[T1]{fontenc}
\usepackage{ae,aecompl}

\usepackage{graphicx}	
\usepackage{amsmath}	
\usepackage{amssymb}	
\usepackage{xcolor}

\newcommand{\uu}[1]{~{\rm #1} }

\title[Cosmic ray heating of ellipticals]{Chaotic cold accretion in giant elliptical galaxies heated by AGN cosmic rays}

\author[Wang et al.]{
Chaoran Wang,$^{1}$\thanks{E-mail: wangcha@umich.edu}
Mateusz Ruszkowski,$^{1}$\thanks{E-mail: mateuszr@umich.edu}
H.-Y. Karen Yang$^{2}$\thanks{E-mail: hsyang@astro.umd.edu}
\\
$^{1}$ Department of Astronomy, University of Michigan, 1085 S. University Avenue, 311 West Hall, Ann Arbor, MI 48109, USA\\
$^{2}$ Department of Astronomy, University of Maryland, College Park, MD 20742, USA}

\date{Accepted XXX. Received YYY; in original form ZZZ}

\pubyear{2019}

\newcommand{\kin}{KINETIC~}
\newcommand{\kmag}{K-MAG~}
\newcommand{\crsh}{CRSH~}
\newcommand{\crcp}{CRCP~}

\begin{document}
\label{firstpage}
\pagerange{\pageref{firstpage}--\pageref{lastpage}}
\maketitle

\begin{abstract}
Black hole feedback plays a central role in shaping the circumgalactic medium (CGM) of elliptical galaxies. We systematically study the impact of plasma physics on the evolution of ellipticals by performing three-dimensional non-ideal magneto-hydrodynamic simulations of the interactions of active galactic nucleus (AGN) jets with the CGM including magnetic fields, and cosmic rays (CRs) and their transport processes. We find that the physics of feedback operating on large galactic scales depends very sensitively on plasma physics operating on small scales. Specifically, we demonstrate that: (i) in the purely hydrodynamical case, the AGN jets initially maintain the atmospheres in global thermal balance. However, local thermal instability generically leads to the formation of massive cold disks in the vicinity of the central black hole in disagreement with observations; (ii) including weak magnetic fields prevents the formation of the disks because local B-field amplification in the precipitating cold gas leads to strong magnetic breaking, which quickly extracts angular momentum from the accreting clouds. The magnetic fields transform the cold clouds into narrow filaments that do not fall ballistically; (iii) when plasma composition in the AGN jets is dominated by CRs, and CR transport is neglected, the atmospheres exhibit cooling catastrophes due to inefficient heat transfer from the AGN to CGM despite Coulomb/hadronic CR losses being present; (iv) including CR streaming and heating restores agreement with the observations, i.e., cooling catastrophes are prevented and massive cold central disks do not form. The AGN power is reduced as its energy is utilized efficiently.
\end{abstract}


\section{Introduction}
In the centers of galaxy clusters, groups, and giant elliptical galaxies, radiative cooling time of the hot gaseous halos can be much shorter than the Hubble time. When this is the case, and in the absence of  heating sources, a classic cooling flow will form \citep{1994ARA&A..32..277F} leading to 
high star formation rates (SFRs). However, the observed SFRs are at least one order of magnitude lower than that predicted by the pure cooling flow \citep[e.g.,][]{2008ApJ...681.1035O}. 
It has been widely accepted that the feedback from active galactic nuclei (AGN) is the most plausible solution to the cooling flow problem. The energetic AGN feedback is able to maintain the overall thermal equilibrium of hot gaseous halos and prevents excessive gas cooling and star formation in galaxy clusters, groups, and giant elliptical galaxies \citep[see][for a review]{2007ARA&A..45..117M}. %
Most AGN found in the centers of these massive systems at low redshift are associated with bipolar jet outflows launched by the central supermassive black holes (SMBHs). The AGN jet inflates radio-emitting bubbles that generate X-ray cavities in the hot halo. How exactly the jets and bubbles transfer energy to the hot halo remains unclear. Various possibilities have been proposed and extensively studied in the literature, including:
mixing of jet-inflated bubble filled with ultra-hot gas \citep[e.g.,][]{2002Natur.418..301B, 2012MNRAS.427.1482G, 2014MNRAS.445.4161H, 2016MNRAS.455.2139H, 2017ApJ...845...91H,Yang2016}; dissipation of jet-induced sound waves and weak shocks \citep[e.g.,][]{Fabian2003,Ruszkowski2004a, Ruszkowski2004b,Fabian2005, 2006MNRAS.366..417F, 2017MNRAS.464L...1F, 2008MNRAS.390L..93S, 2018ApJ...858....5Z, Fabian2003, Fabian2005, 2017MNRAS.464L...1F, 2007ApJ...665.1057F, 2015ApJ...805..112R, 2016MNRAS.461.1548B, Yang2016, Li2017,2019MNRAS.483.2465M}, and turbulent dissipation of gas motions \citep[e.g.,][]{2010ApJ...710..743D, 2014Natur.515...85Z}; and heating by cosmic rays (CRs) escaping from the jet-inflatd bubbles \citep[e.g.,][]{GuoOh2008,2011A&A...527A..99E,Ruszkowski2008,2013ApJ...767...87W,2013MNRAS.428..599F,2013MNRAS.432.1434F, 2017MNRAS.467.1449J, 2017MNRAS.467.1478J, Ruszkowski2017, 2018MNRAS.481.2878E}. However, no consensus has been reached and the dominant mechanism(s) is(are) uncertain. 

Despite adopting various numerical methods, simulations of AGN feedback and cooling flow problem often reveal formation of a massive long-lived cold disk in the vicinity of the central SMBH \citep[e.g.,][]{2014ApJ...789...54L, 2017MNRAS.468..751E, 2018arXiv181001857Q, 2019MNRAS.482.3576W}. However, the observed molecular gas rarely appears to be rotationally supported in either elliptical galaxies \citep{Young2011} or central cluster galaxies \citep{2019arXiv190209227R}. Moreover, these simulations produce overly massive disks compared to the few disks found in the observations. The formation of such massive disks poses further problems for the feeding of the central SMBH and AGN feedback. In our previous work \citep{2019MNRAS.482.3576W}, the fueling of large amounts of cold gas from the disk breaks the self-regularity of the AGN feedback, thus causing the central region of the halo to overheat. This results in an isentropic core, which is inconsistent with the observations.

Radio observations show that the jet-inflated bubbles emit synchrotron radiation, indicating the presence of relativistic electrons. However, it has been shown that the emitting electrons plus magnetic field have less pressure than required to maintain pressure balance between the bubble interior and the ambient hot halo gas \citep{DunnFabian2004}. Therefore, the non-radiating component has to be present inside the bubbles to provide the required pressure, e.g., ultra-hot gas, magnetic fields, and CR protons. However, the relative contribution of each of these non-thermal components is not well constrained by the observations.

CR-dominated AGN jet can potentially solve the disk problem mentioned above.  CRs heat the halo gas via Coulomb, hadronic interactions \citep{yoasthull2013}, and streaming instability \citep[][for a review]{zweibel2013} as they are transported out from the jet-inflated bubbles. Streaming instability allows CRs to scatter on the self-excited Alfv{\'e}n waves and to stream down their pressure gradient along the magnetic field at or above the Alfv{\'e}n speed. CR streaming down the pressure gradient along the B-field results in CRs doing work on the gas and effectively heating it. This additional mode of heating could help to heat the has at larger distances from the center and eliminate catastrophic cooling and cold disk formation at late times. However, recent simulations of a single short-duration jet in the galaxy cluster \citep{2019ApJ...871....6Y} demonstrate that CR-dominated jets can efficiently uplift the hot halo due to high buoyancy of the CR fluid filling the bubbles and larger cross-sections of the raising bubbles. This could remove gas with the shortest cooling time near the cluster center and potentially limit the formation of the disk in the long term. Interestingly, while more efficient heating in long term is expected, \citet{2019ApJ...871....6Y} show that the CR-dominated jet makes the intracluster medium (ICM) more prone to the development of thermal instabilities in the center due to inefficient mixing and heating on short timescales. 

Motivated by the above findings, in this paper we investigate how CR feedback affects the long-term evolution of the hot halo. Using magneto-hydrodynamic (MHD) simulations, we show that when CR transport and heating due to the streaming instability are included, CR-dominated jet can maintain global thermal equilibrium of the hot halo in elliptical galaxies and, at the same time, prevent the formation of the massive cold disks.

\begin{figure*}
  \begin{center}
    \leavevmode
    \includegraphics[width=\textwidth]{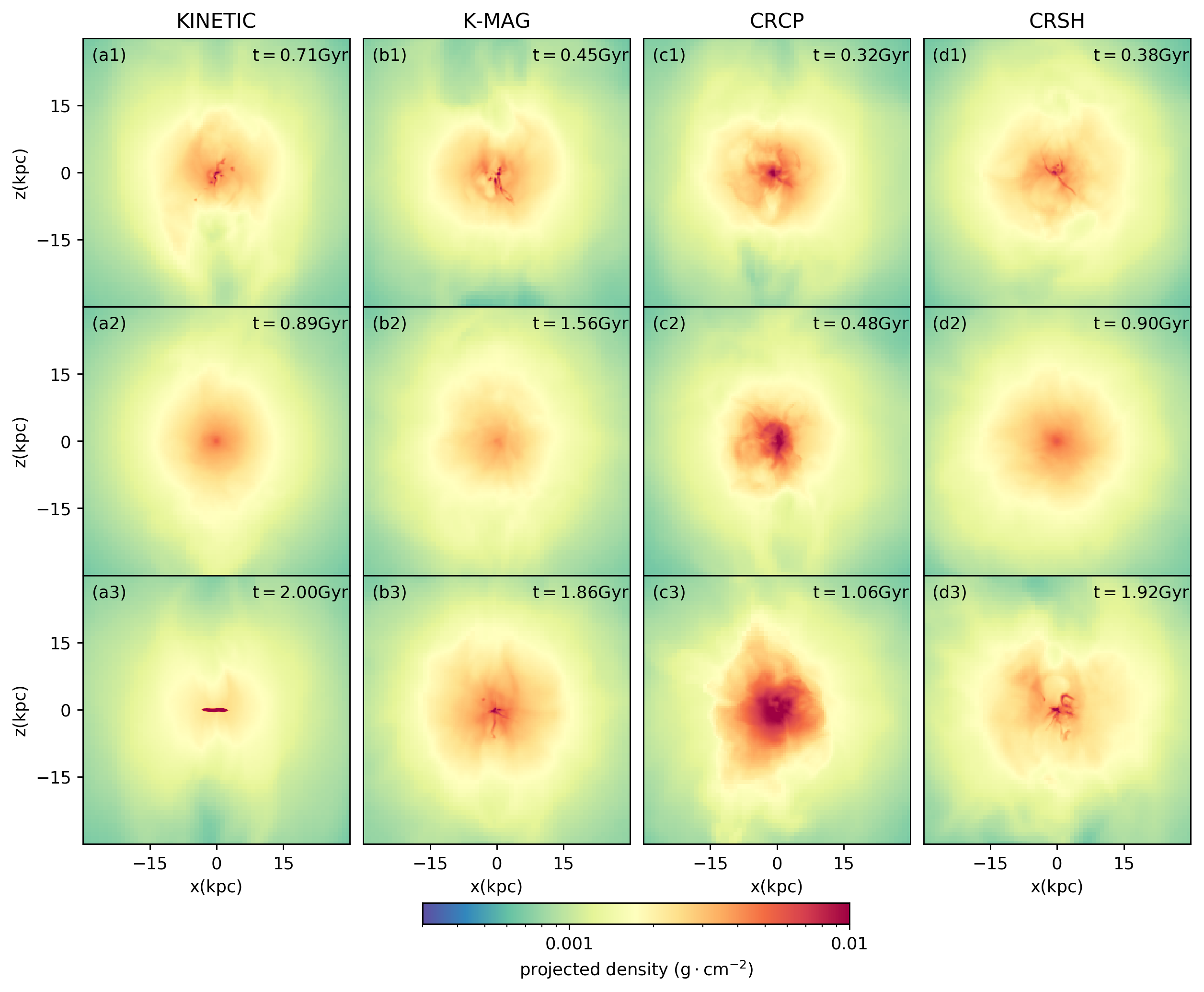}  
       \caption[]{Columns from left to right: projected gas density of KINETIC, K-MAG, CRCP, and CRSH runs. Within each column the slices are plotted in chronological order from top to bottom. The density is projected along the $y-$axis within the central 60~kpc-wide cube. The jet is along the $z-$axis. }
     \label{fig:1}
  \end{center}
\end{figure*}

\begin{figure*}
  \begin{center}
    \leavevmode
    \includegraphics[width=0.8\textwidth]{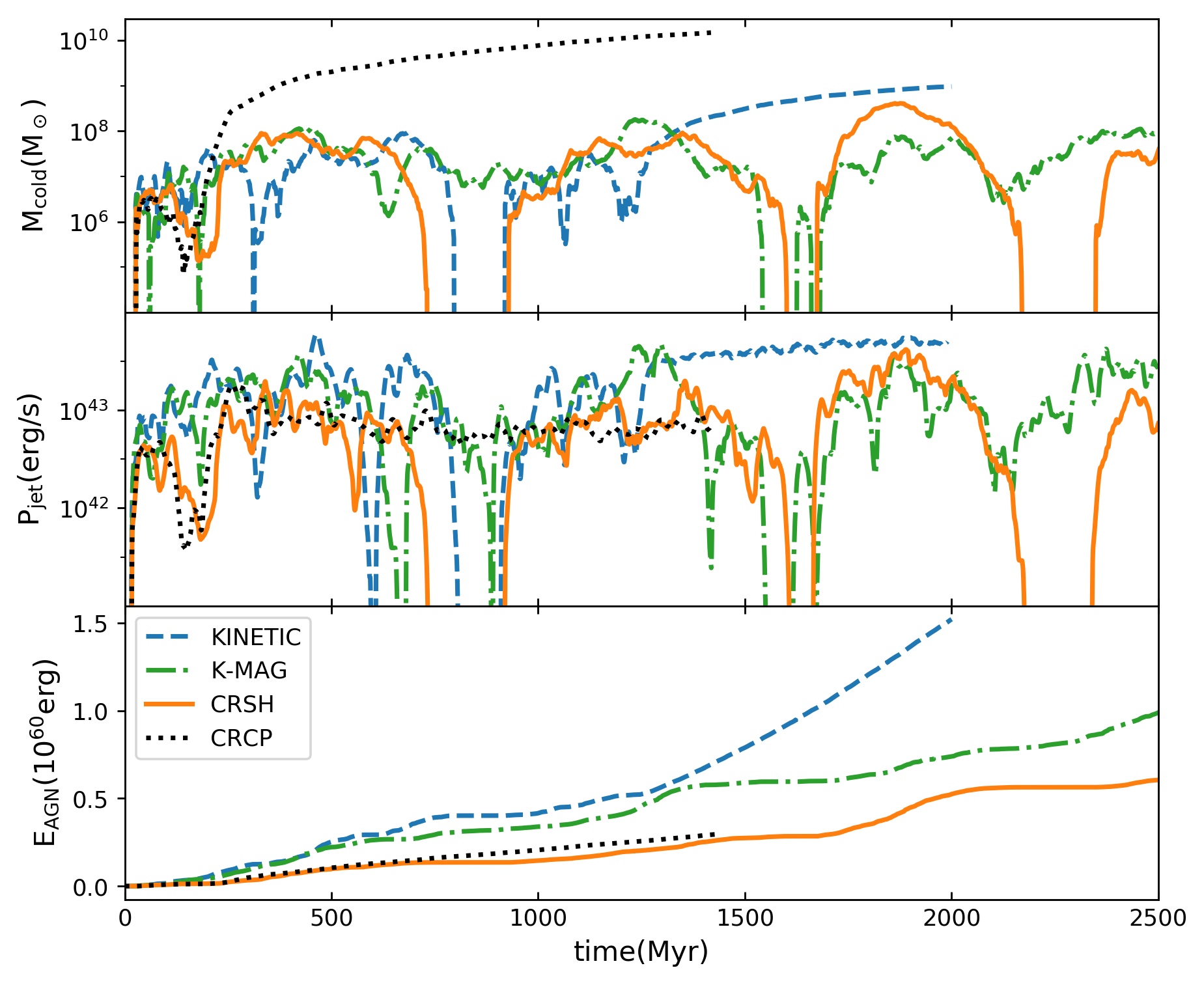}  
       \caption[]{Time evolution of cold gas mass (top), AGN jet power (middle) and cumulative AGN energy (bottom) of KINETIC (blue dashed lines), K-MAG (green dash-dotted lines), CRCP run (black dotted lines), and CRSH (orange solid lines). The jet powers are averaged using a $20\uu{Myr}$-wide moving window. }
     \label{fig:2}
  \end{center}
\end{figure*}

\section{Methodology}
\begin{table}
\caption{List of Simulations}
\centering
 \begin{tabular}{ccccccccc} 
\hline
name  & magnetic field & $f_{\rm cr}$ &  CR streaming \\
 \hline
 \kin            & no             &  0           &  no  \\
 \kmag           & yes            &  0           &  no  \\
 \crcp           & yes            &  0.8         &  no  \\
 \crsh           & yes            &  0.8         &  yes \\
 \hline
 \end{tabular}
\end{table}

We perform three-dimensional MHD simulations using the FLASH code \citep{2000ApJS..131..273F, dubey2008introduction} with the directional unsplit staggered mesh MHD solver \citep{Lee09, Lee13}. The simulation domain is a statically-refined cubic box with of $L_{\rm box}=250~\rm kpc$. The domain is refined by a set of nested cubic regions with increasing refinement levels. The entire region is refined by $(64)^3$ root grids and the central $L_{\rm box}/2^{n}$ wide regions have $n$ additional nested levels of refinement ($n=1,...,4$) with the refinement level increasing with the decreasing distance from the domain center. Therefore, the smallest cell is $\Delta x_{\rm min}=L_{\rm box}/64/2^4\approx0.244~\rm kpc$ wide and the highest refinement region is $L_{\rm box}/2^{4}\approx16\uu{kpc}$ wide. We adopt the diode boundary conditions. That is, the gas can only flow out and all variables have zero gradient at the computational box boundary. 

We follow our previous work \citep{2019MNRAS.482.3576W} in setting up the initial conditions. The density and temperature profiles of the hot gas are initially in hydrostatic equilibrium. The gravitational potential includes contributions from stars, dark matter, and the central SMBH. Our initial conditions agree with NGC 5044, which is observed to have extended multiphase gas in the center (\citealt{Werner2014}; as explained below, our simulations develop cold phase component over time due to local thermal instability). Following \citet{Ruszkowski2007}, we include tangled magnetic fields with the following power spectrum: \begin{equation}
 \label{eq:mag}
    B_k \propto k^{-11/6} \exp{\left[-\left(\frac{k}{k_0}\right)^{4}\right] },
\end{equation}
where $k_0=10^2(2\pi/L_{\rm box})$. In order to generate magnetic fields characterized by the above spectrum, we first calculate real-space magnetic fields by inversely Fourier transforming the above power spectrum. We then renormalize the average field strength such that the plasma $\beta\sim 100$. We then Fourier transform the B-fields, clean B-field divergence in Fourier space and then perform inverse Fourier transform to obtain real-space B-fields. We repeat this procedure until the magnetic fields become divergence-free.

The key physical processes that we consider are radiative cooling, feedback from old stars, CR transport and heating, and AGN feedback. For radiative cooling, we use the tabulated cooling functions from \cite{SutherlandDopita93} assuming one solar metallicity \citep{2003ApJ...595..151B, 2009PASJ...61S.337K}. For old star feedback, we use the model described in \cite{2019MNRAS.482.3576W}. In short, we consider the  mass loss from evolved stars and  thermal energy injection from type Ia supernovae. The ejecta from the evolved stars are assumed to be thermalized by the random stellar motions and the energy and mass are injected at a constant rate.
 
In the following sections, we describe the included CR physics (section \ref{method:mag}) and the AGN feedback model (section \ref{method:agn}).

\subsection{CR physics}\label{method:mag}
The CRs are modelled as a relativistic fluid with adiabatic index $\gamma_{\rm cr}=4/3$. We follow the evolution of the CR pressure $p_{\rm cr}$ (or energy density $e_{\rm cr}=p_{\rm cr}/(\gamma_{\rm cr} -1)$)  as a function of space and time. The CR implementation includes the following processes: 
(i) advection with the gas, 
(ii) transport of CRs along the magnetic field lines due to Alfv{\'e}nic streaming (CR streaming/drift speed is determined by the streaming instability) and the associated energy transfer from CRs to the gas, and (iii) heating of gas by the CRs via Coulomb and hadronic interactions. We solve the MHD equations in the following form:

\begin{equation}
\frac{\partial \rho}{\partial t} + \nabla \cdot(\rho{\bf u_g}) = 0    
\end{equation}
\begin{equation}
    \frac{\partial (\rho {\bf u_g})}{\partial t} + \nabla\cdot\left(\rho {\bf u_g u_g} - \frac{\bf BB}{4\pi}\right) + \nabla p_{\rm tot}= \rho {\bf g}
\end{equation}
\begin{multline}\label{eq:ener}
\frac{\partial e_{\rm tot} }{\partial t}+ \nabla \cdot\left[(e_{\rm tot} + p_{\rm tot}){\bf u_g} - \frac{({\bf B\cdot u_g}){\bf B} }{4\pi} +  {\bf F}_{\rm stream} \right] = \rho {\bf u_g\cdot g}  \\ - C_{\rm rad} - C_{\rm cr, net} 
\end{multline}
\begin{multline}\label{eq:ecr}
\frac{\partial e_{\rm cr}}{\partial t} 
+ \nabla \cdot(e_{\rm cr}{\bf u_g} + {\bf F}_{\rm stream}) = -p_{\rm cr}\nabla\cdot{\bf u_g} \\   -C_{\rm cr, C} - C_{\rm cr, h} - C_{\rm cr, s},
\end{multline}
where $\rho$ is the gas density; $\bf u_g$ is the gas velocity; $\bf B$ is the magnetic field; and $\bf g$ is the gravitational acceleration. The total pressure, $p_{\rm tot}$ is the sum of the gas thermal pressure, magnetic pressure and the CR pressure,
$p_{\rm tot} = p + B^{2}/8\pi + p_{\rm cr}$; and the total energy density $e_{\rm tot}$ is the sum of the gas internal energy, kinetic energy, magnetic energy, and the CR energy $e_{\rm tot} = p/(\gamma_{\rm gas} -1) + \rho u_g^{2}/2 + B^{2}/8\pi + e_{\rm cr}$, where $\gamma_{\rm gas}=5/3$ is the adiabatic index of the ideal gas. The three terms on the right hand side of Eq.\ref{eq:ecr}, $C_{\rm cr, C}$, $C_{\rm cr, h}$, and $C_{\rm cr, s}$, account for the CR fluid energy losses due to Coulomb, hadronic interactions, and streaming heating, respectively.
$C_{\rm cr, C}$ and $C_{\rm cr, h}$ are given by \citep{yoasthull2013}:
\begin{equation}
    C_{\rm cr, C} = 4.93 \times 10^{-19}
    \left(\frac{n_{\rm cr}}{\rm cm^{-3}}\right)
    \left(\frac{n_{\rm e}}{\rm cm^{-3}}\right)
    {\rm erg\cdot cm^{-3}s^{-1}};
\end{equation}
\begin{equation}
    C_{\rm cr, h} = 8.56 \times 10^{-19}\left(\frac{\gamma}{3}\right)^{1.28} \left(\frac{n_{\rm cr}}{\rm cm^{-3}}\right)
    \left(\frac{n_{\rm p}}{\rm cm^{-3}}\right) {\rm erg\cdot cm^{-3}s^{-1}},
\end{equation}
where $\gamma=3$ is the average Lorentz factor of CRs, $n_{\rm p}=\rho/(\mu_{\rm p} m_{\rm p})$ is the gas proton number density, and $n_{\rm e}=\rho/(\mu_{\rm e} m_{\rm p})$ is the gas electron number density. $n_{\rm cr}$ is the CR number density which is:
\begin{equation}
n_{\rm cr} = \frac{n-4}{n-3} \frac{e_{\rm cr}}{E_{\rm min}},
\end{equation}
where $E_{\rm min}=1\rm GeV$ is the low energy cut-off of CR energy, and $n=4.5$ is the slope of CR momentum distribution. In the self-confined picture, CRs stream with respect to the gas and get scattered on self-excited MHD waves. This leads to the transfer of CR energy to the gas. The heating rate is \citep{wiener2013}:
\begin{equation}
    C_{\rm cr, s} = |{\bf u_A'}\cdot \nabla p_{\rm cr}|, 
\end{equation}
where $\bf u_A'$ is the modified Alfv{\'e}n velocity. We assume: 
\begin{equation}
{\bf u'_A} =  \left\{\begin{array}{lr}
        {\rm min}({\bf u_A}, 200~{\rm km/s}), & r > 1~{\rm kpc} \\
        0, & r \leqslant 1~{\rm kpc} \\ 
        \end{array}\right.
\end{equation}
The size of the central region where the streaming heating is suppressed is comparable to the size of the base of the AGN jet. We treat the central 1~kpc region as a subgrid region and switch off streaming heating there to avoid artificial heating. In this approximation the streaming velocity cannot exceed than 200~km/s, which helps to save computational cost. We verified that this constraint limits the heating in only in a very small fraction of cells and does not affect the conclusions presented below.

The terms on the right hand side of Eq.\ref{eq:ener}, $C_{\rm rad}$ and $C_{\rm cr, net}$, represent the radiative cooling losses of the gas and the net total energy loss of the composite gas plus CR fluid, respectively. As during hadronic collisions 5/6 of the CR energy escapes from the gas due to gamma-ray and neutrino emission, the net loss of the energy is
\begin{equation}
C_{\rm cr, net} = \frac{5}{6}C_{\rm cr, h}, 
\end{equation}
i.e., 1/6 of the CR hadronic loss rate is used to heat the thermal gas, and the remaining 5/6 are permanently lost from the system.

The CR streaming flux ${\bf F}_{\rm{stream}}$ in Eq.\ref{eq:ener} and Eq.\ref{eq:ecr} is ${\bf F}_{\rm{stream}} = -(e_{\rm cr}+p_{\rm cr}){\bf u'_A}~{\rm sgn}\left({\bf B}  \cdot \nabla e_{\rm cr}\right)$. To avoid infinitely fast variations of $\nabla\cdot {\bf F}_{\rm{stream}}$ at the local extrema of $\nabla e_{\rm cr}$, we follow the method of \citet{Sharma2009} and regularize the streaming flux by a scale height $h_c=10\uu{kpc}$: 
\begin{equation}
    {\mathbf F}_{\rm stream} = -(e_{\rm cr}+p_{\rm cr} ) \mathbf{u'_A}~{\rm tanh}\left( h_c \frac{\bf B}{|{\bf B}|}\cdot \frac{\nabla  e_{\rm cr}}{e_{\rm cr}}  \right).
\end{equation}

\begin{figure*}
  \begin{center}
    \leavevmode
    \includegraphics[width=0.9\textwidth]{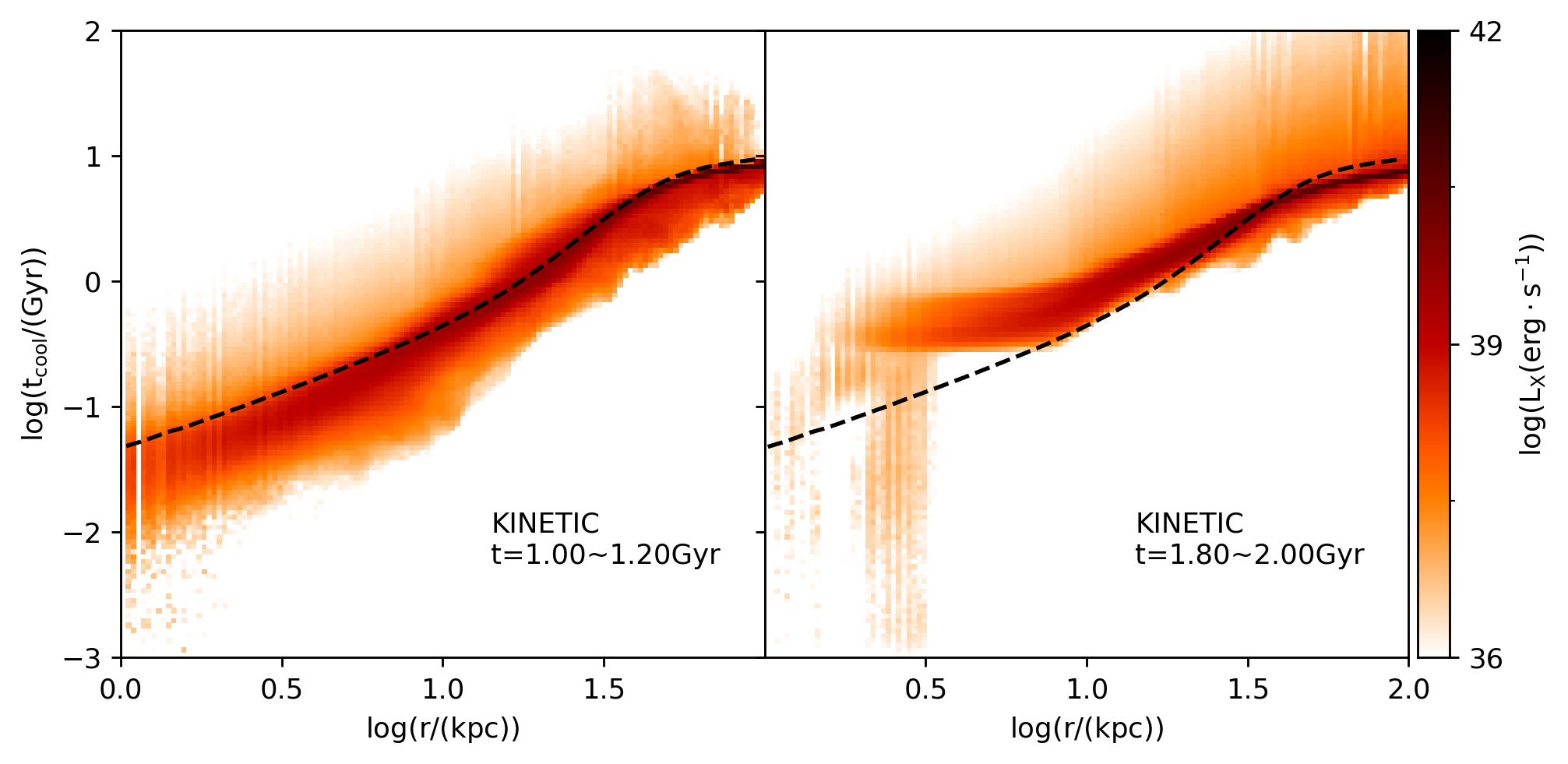}  
       \caption[]{The radial distributions of gas cooling time in KINETIC run
       averaged over two periods of time.  
       Left:  $t=1\sim1.2\uu{Gyr}$; right $t=1.8\sim2.0\uu{Gyr}$. For each panel, the average is calculated by stacking the cooling time distributions of all simulation outputs within the corresponding period.  
       Color shows the gas X-ray luminosity. The dashed lines represent the initial cooling time profile.}
     \label{fig:3}
  \end{center}
\end{figure*}

\subsection{AGN feedback}\label{method:agn}
For the fueling of the AGN, we adopt the cold mode accretion model, which has been used in many numerical studies in the literature (e.g., \citealt{GRS,GRO,2014ApJ...789...54L,li2014,Yang2016}). The cold gas ($T<10^5\uu{K}$) within the central $r<0.5\uu{kpc}$ region is accreted at the rate $\dot{M}_{\rm acc}=M_{\rm in,c} / 5\uu{Myr}$, where $M_{\rm in,c}$ is the mass of cold gas within $r<0.5~{\rm kpc}$. The accretion-powered AGN feedback injects gas and CRs. The gas feedback is purely kinetic with kinetic power ($\dot{E}_{\rm k}$) given by 
\begin{align}
    \dot{E}_{\rm k} =& \frac{1}{2}\eta \dot{M}_{\rm acc} v_{\rm jet}^2 = \epsilon \eta (1-f_{\rm cr}) \dot{M}_{\rm acc} c^2; 
\end{align}
and the CR feedback power ($\dot{E}_{\rm cr}$) given by
\begin{align}
    \dot{E}_{\rm cr} =& \epsilon\eta f_{\rm cr} \dot{M}_{\rm acc} c^2,
\end{align}
where $c$ is the speed of light, $\epsilon$ is the feedback efficiency, $\eta$ is the mass loading factor and $f_{\rm cr}$ is the CR fraction of the AGN injected energy. We adopted $\epsilon=3\times10^{-4}$ and $\eta=1$ for all simulations in this work and $f_{\rm cr}=0.8$ for all simulations that included CRs. The AGN ejecta is launched via bipolar jets. The jet launching base is the innermost 8 cells and the jet precesses around the $z-$axis with the precession angle of 10 degrees and period of $10\uu{Myr}$.

\section{Results}
We now systematically discuss the results from our simulations. In this section we comment on simulations with progressively more complex physics of the CGM plasma, i.e., we begin with a purely hydrodynamical run (KINETIC run), then discuss its generalization to include magnetic fields (K-MAG run) and CRs including Coulomb and hadronic/pionic losses (CRCP run), and finally comment on the run that extends the above physics to include CR transport via the streaming instability and the associated with it CR heating of the CGM (CRSH run). Key parameters distinguishing these runs are shown in Table 1, where $f_{\rm cr}$ denotes the fraction of the CR energy in the AGN jet.

\subsection{AGN feedback in the hydrodynamical case}
In the purely hydrodynamical AGN feedback case, the atmosphere goes through cycles of cold gas precipitation and AGN outbursts in the first $1.2\uu{Gyr}$. At the beginning of each cycle, AGN outburst is triggered by either (i) central gas cooling due to short central cooling time, or (ii) infall of cold gas that condenses out of previously uplifted low entropy gas. The first cycle is initiated by the short central cooling time. The AGN jets perturb the halo triggering thermal instabilities. Consequently, cold gas condenses out of the halo and forms extended structures (see panel a1 in Fig.\ref{fig:1}). Some of the cold clumps fall toward the center, swirl around the central SMBH and eventually get accreted by it. A fraction of cold clumps find themselves in the path of the AGN jet, get fragmented by it, and get either ablated or accreted by the SMBH. Cold gas condensation ends when the AGN heating makes the halo thermally stable. When all cold gas gets consumed or destroyed, the AGN shuts down and the halo returns to the quiescent state (see panel a2 in Fig.\ref{fig:1}) until the next cycle begins. During the precipitation cycles, the cold gas mass fluctuates between $\sim 10^{6}$ and $\sim 10^{8}\uu{M_{\odot}}$, and the AGN jet power typically ranges between $10^{42.5}\sim10^{43.5}\uu{erg\cdot s^{-1}}$ (see top and middle panels in Fig.\ref{fig:2}). This cyclic evolution is a natural result of the self-regulated AGN feedback. Many previous studies of cold mode accretion AGN feedback in giant elliptical galaxies or galaxy clusters led to similar conclusions with regard to the behavior of cold gas mass and jet power fluctuations. The cyclic evolution only lasts for the first $1.2\uu{Gyr}$, after which the simulation goes into the unphysical {\it disk phase} described below. 

\subsubsection{Formation of a massive cold disk in the vicinity of the SMBH}\label{ssec:disk}
In the third precipitation cycle accreting cold gas forms a rotationally-supported disk at $t\approx1.2\uu{Gyr}$. Although the formation of such disks has been seen in many simulations \citep[e.g.,][]{2014ApJ...789...54L, 2017MNRAS.468..751E, 2018arXiv181001857Q, 2019MNRAS.482.3576W}, we consider the evolution after the disk formation to be unphysical due to the following two reasons.
\begin{enumerate}
    \item The disk quickly becomes too large and too massive. Since the onset of disk formation, the cold gas mass in the disk increases monotonically as shown in the top panel of Fig.\ref{fig:2}. At $t=2\uu{Gyr}$, the disk grows to $3\uu{kpc}$ in radius (third row of the left column in Fig.\ref{fig:1}) and $10^9\uu{M_\odot}$ in mass. Such massive cold disk is not found in NGC5044 nor in most of the observed elliptical galaxies \citep{Young2011}. 
    \item The feeding of the AGN from the disk leads to the overheating of the central few kpc region while hot gas continues to cool directly onto the disk. Fig.\ref{fig:3} shows the average radial distribution of gas cooling time, $t_{\rm cool}(r)$,  over two $200\uu{Myr}$-long intervals -- one just before the disk formation, $t=1\sim1.2\uu{Gyr}$ (left panel) and the other one $600\uu{Myr}$ later, $t=1.8\sim2\uu{Gyr}$ (right panel);  the dashed lines in both panels represent the initial $t_{\rm cool}(r)$. Before the formation of the disk, the system is self-regulated, and the gas cooling and AGN heating are overall in balance with each other. Therefore, $t_{\rm cool}(r)$ is scattered approximately symmetrically around the initial values. In this phase, the cold gas is accreted onto the SMBH and the cold gas is removed from the central regions. However, in later phase, the rate of cold gas consumption by the SMBH and the jet power reach a plateau. This happens because a centrifugally supported disk is formed and the angular momentum is not sufficiently quickly removed from it. Furthermore, because the disk rotates about the jet axis, the AGN outflow cannot intercept the cold gas and is unable to reheat it (this type of reheating is present in the simulations that exhibit cyclic AGN behavior). Therefore, the AGN consistently heats the halo while the disk continues to grow in mass as the hot phase cools and adds mass to it. As shown in the right panel of Fig.\ref{fig:3}, while $t_{\rm cool}(r)$ systematically increases away from the initial condition due to the AGN heating in the $3\uu{kpc}<r<10\uu{kpc}$ region, it cascades down for $r\lesssim3\uu{kpc}$ because the gas directly cools onto the disk. Therefore, even though the AGN overheats the halo, it fails to stop the fast cooling of the gas. The whole system is no longer self-regulated and its thermal properties are systematically changed compared to the initial condition.
\end{enumerate}

\subsection{AGN feedback in the magneto-hydrodynamical case}

We now discuss the impact of magnetic fields on the purely hydrodynamical simulations of the AGN feedback described above. Similarly to the \kin case, in the \kmag case the atmosphere and the AGN undergo cyclic variations. In the \kmag case thermal instability is also present and cold gas clouds decouple from the hot atmosphere. However, unlike in the \kin case, massive cold disk surrounding the SMBH does not form
(c.f. last row of the first and second column in Fig. 1). Moreover, the atmosphere remains in a state of global thermodynamic equilibrium without overheating in the time-averaged sense throughout the simulation time rather than exhibiting overheating that occurs in the \kin case at about 1.2 Gyr. As we argue below, both of these differences can be attributed to the impact of magnetic tension acting on the clouds and magnetic breaking that leads to more efficient extraction of angular momentum and accretion of the cold clumps onto the central SMBH. 

\begin{figure*}
  \begin{center}
    \leavevmode
    \includegraphics[width=0.9\textwidth]{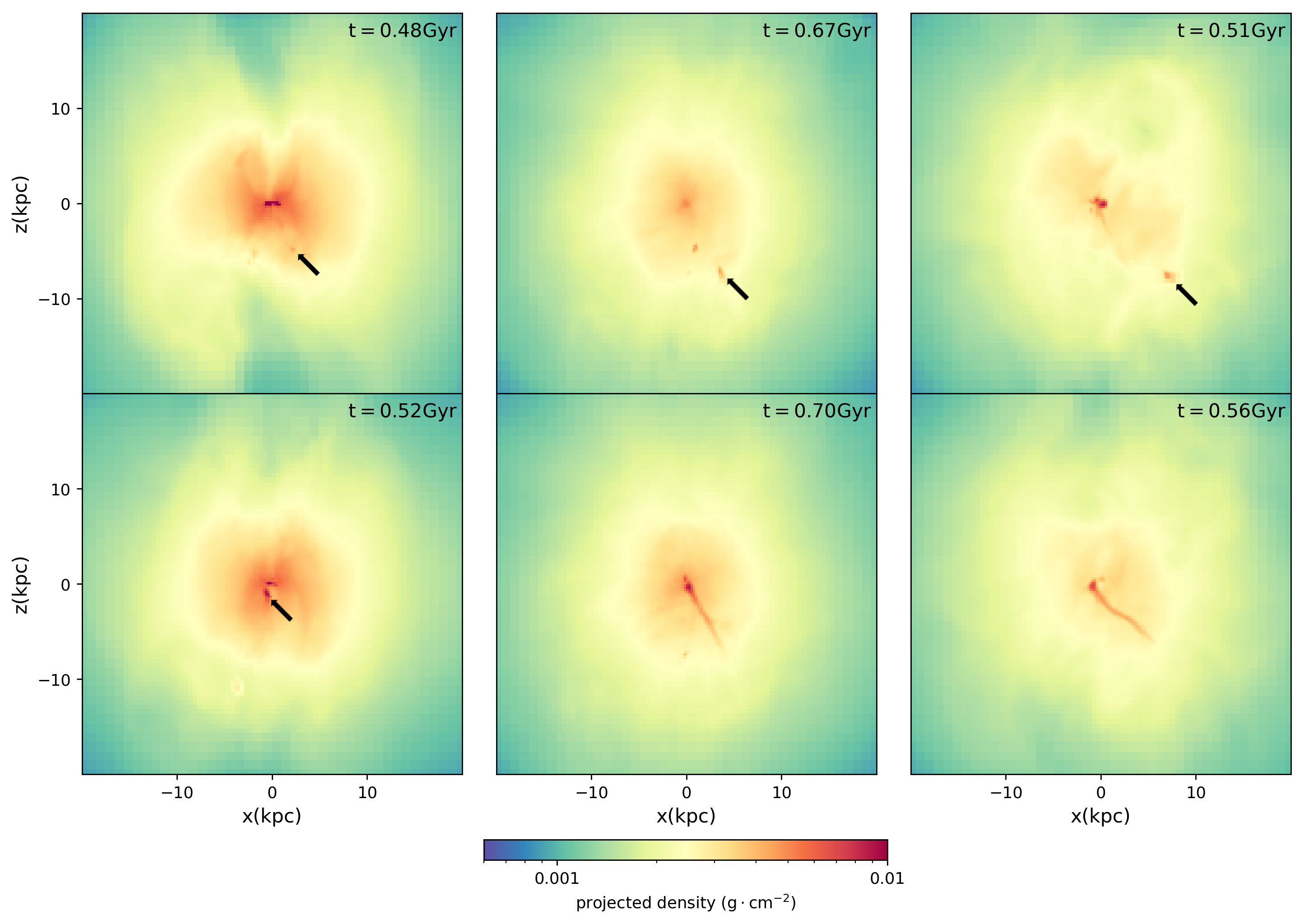}  
       \caption[]{Infall of an isolated cold clump in (from left to right column) KINETIC, KMAG, and CRSH run. In each column, the top row is the projected density snapshot at the time when the condensed cold clump starts to fall in and the bottom row corresponds to the epoch when the head of the clump reaches the center. Arrows point to the locations of the cold clouds when they are compact. }
     \label{fig:filaments}
  \end{center}
\end{figure*}

\subsubsection{Morphology of the precipitating clouds}
In order to unravel the reason for the absence of cold disks in the \kmag case, we now turn our attention to the morphology of the clouds. To this end, we compare how cold clumps evolve in the \kin and \kmag runs and analyze one event of the infall of an isolated cold clump in each run. Fig.\ref{fig:filaments} shows snapshots of the projected density at the beginning of the infall of isolated clumps (top row) and at the time when the cold clumps reach the center (bottom row). Arrows point to the locations of the cold clouds when they are compact. 
While in the \kin case (left column) the cloud appears compact throughout its infall, in the \kmag case the cloud is compact in the beginning and then becomes extremely filamentary as it reaches the center. The filaments are approximately aligned with the radial direction, which to first order coincides with the infall trajectory (see middle column). This difference in the  morphology of the cold gas clouds between \kin and \kmag cases strongly suggests that magnetic forces may be important. We note in passing that the clouds in the \kin case also become filamentary in those cases where they directly interact with the AGN outflow.

\begin{figure}
  \begin{center}
    \leavevmode
    \includegraphics[width=\columnwidth]{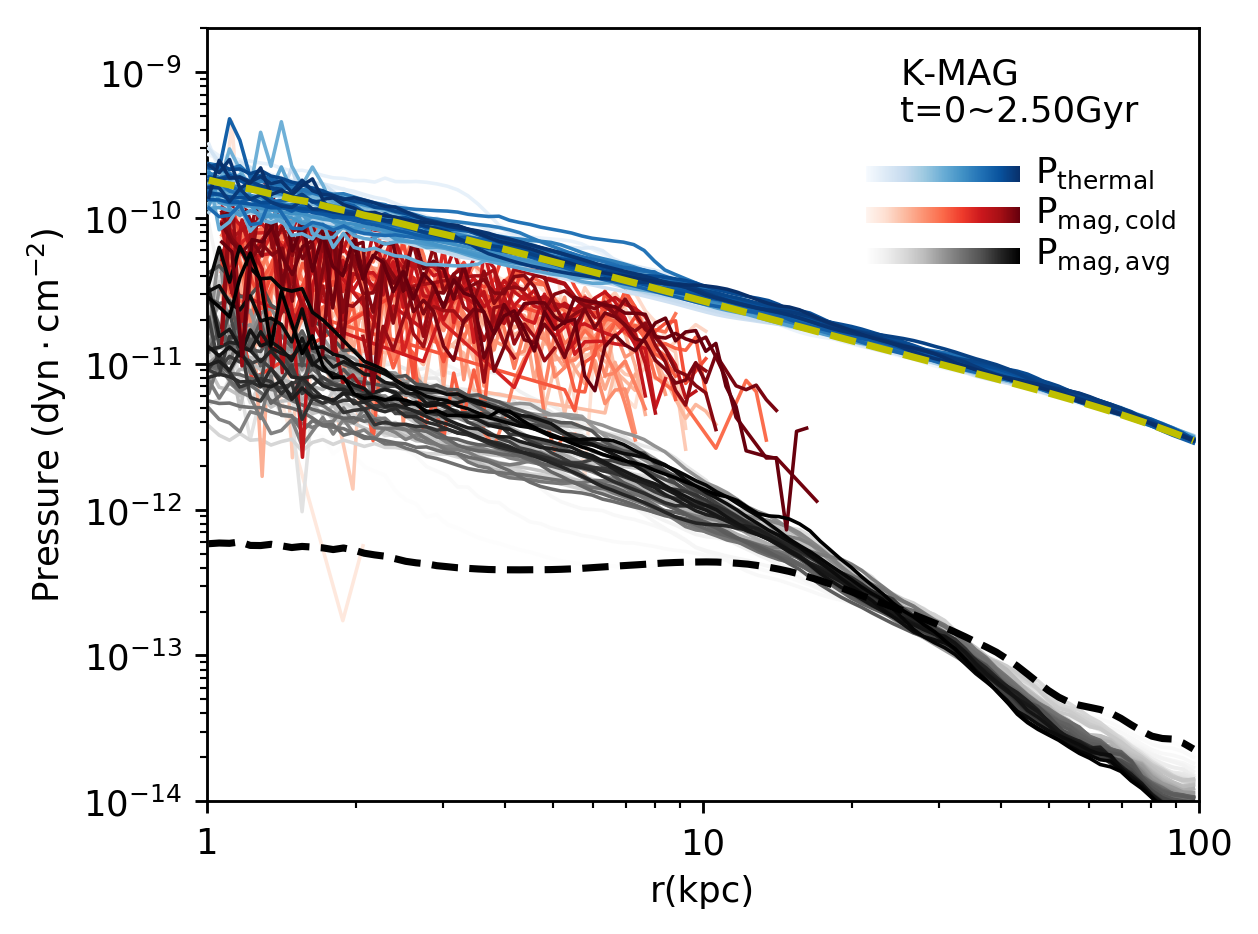}  
       \caption[]{Three sets of pressure profiles: thermal pressure (blue), magnetic pressure in the regions occupied by the cold gas (red), and magnetic pressure averaged in spherical shells (grey). For each set, the profiles are plotted every $50\uu{Myr}$ from light to dark color. For clarity, the initial thermal and magnetic pressure profiles are shown as yellow and black dashed lines, respectively.}
     \label{fig:5}
  \end{center}
\end{figure}

\subsubsection{Impact of magnetic fields on the cold phase morphology}
Although the mean magnetic field in the CGM is dynamically weak (initial plasma $\beta\sim 100$), this does not preclude a possibility that the B-field could be locally enhanced. In Fig.\ref{fig:5} we show the profiles of the thermal pressure (in blue), magnetic pressure averaged in spherical shells (in gray), and the magnetic pressure in cold gas regions ($T<10^{5}$K). In each category, different lines correspond to different simulation times with darker lines corresponding to later times. As expected, the average plasma $\beta\gg 1$. Interestingly, the cold gas phase is spatially correlated with enhanced magnetic field strength. 

\subsubsection{Magnetic breaking and cold gas velocity distribution}
\begin{figure*}
  \begin{center}
    \leavevmode
    \includegraphics[width=1.0\textwidth]{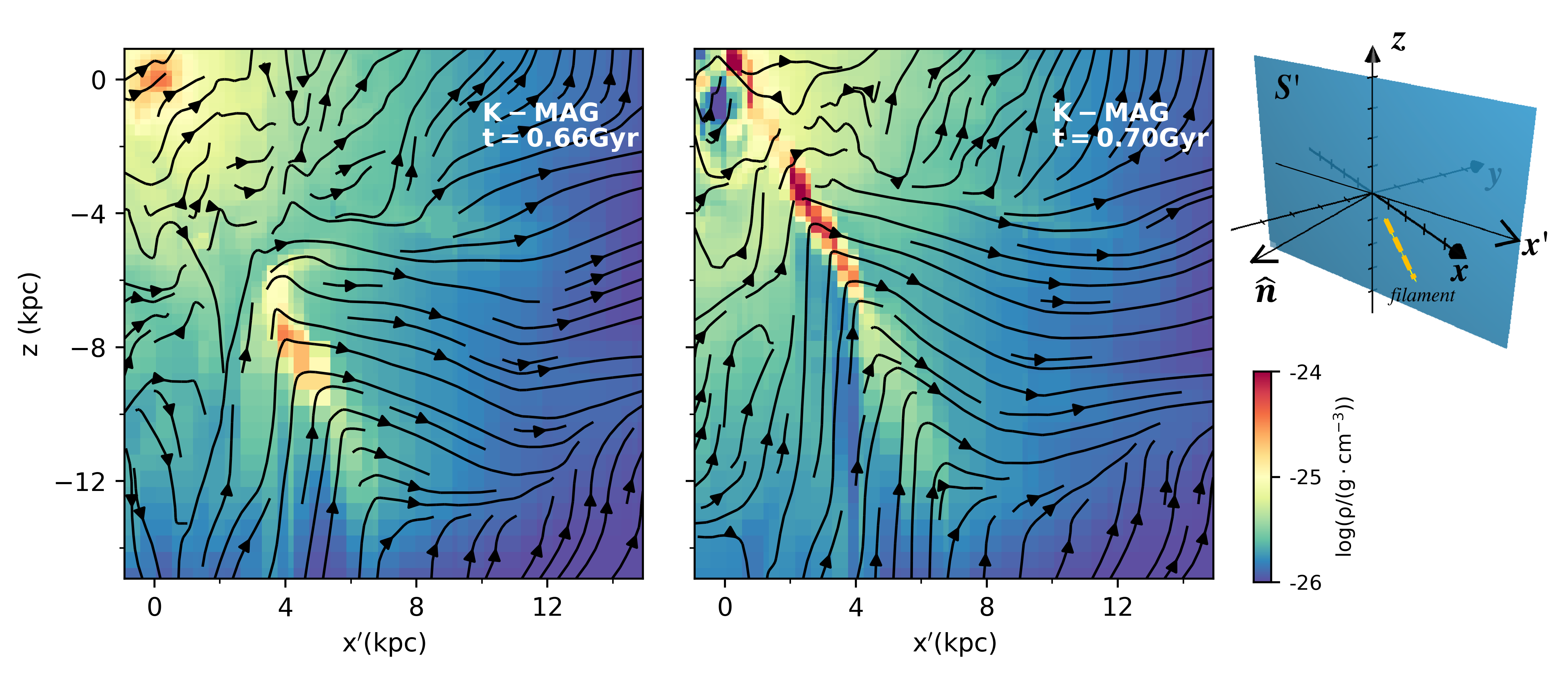}  
       \caption[]{Left and middle panel: distribution of gas density and magnetic field lines on the plane $S'$ at the epoch of the formation of a single filament (left, $t=0.66\uu{Gyr}$) and at the epoch when the filament head reaches the center (middle, $t=0.70\uu{Gyr}$). The gas density and magnetic field lines are shown as color maps and stream lines, respectively. Upper right panel: the position of plane $S'$ relative to the $x, y, z$ axes of the simulation box. The normal vector and the intersection with the $x-y$ plane of $S'$ are labelled as $\hat{n}$ and $x'$, respectively. The plane $S'$ slices the volume through the bulk of the filament, which is schematically shown as the orange dashed line. 
       }
     \label{MagneticTails}
  \end{center}
\end{figure*}
\indent
The local B-field amplification occurs due to flux freezing during the condensation of the cold gas followed by the field amplification during cloud infall. This amplification could occur on the leading edge of the cloud as well as in the filamentary tail. As the clumps fall to the center, the strong magnetic fields are stretched and exert tension force in the direction anti-parallel to the clump velocity direction. This mechanism is clearly shown in Fig.\ref{MagneticTails}, which shows the gas density distribution on a plane containing the cold filament at two different times (see left and middle panels). The spatial orientation of this plane is shown in the upper right panel of Fig.\ref{MagneticTails}. The magnetic field lines are shown as arrows. As the cold clump is falling to the center, it bends and stretches the magnetic field lines. The field lines exert strong tension force on the cold clump. Any gas that is stripped from the cloud is confined to the ordered fields lagging behind the cloud and slides along these fields forming long cold filaments.\\
\begin{figure}
  \begin{center}
    \leavevmode
    \includegraphics[width=1.0\columnwidth]{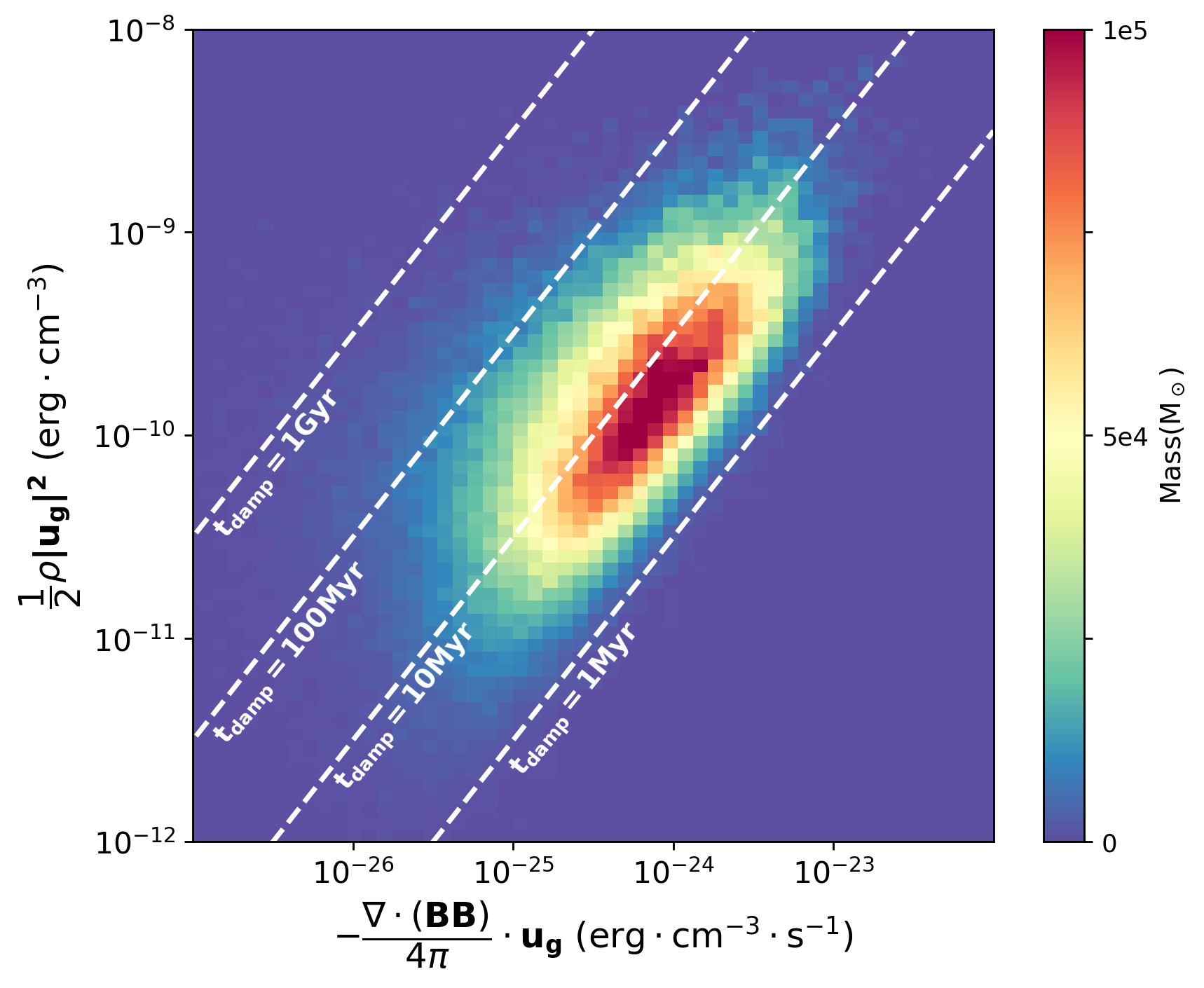}
       \caption[]{Kinetic energy - magnetic tension ``resistence'' 
       phase plot of the cold gas in \kmag run. Color represents the cold gas mass. The distribution is averaged over snapshots  sampled every $2\uu{Myr} $ within $t=0-2.5\uu{Gyr}$. Dashed lines denote the damping time scale $t_{\rm damp}=1, 10, 100\uu{Myr}$ and $1\uu{Gyr}$. The damping timescale is defined as the timescale over which the work done by the magnetic tension force depletes the kinetic energy of the gas.
       }
     \label{fig:tension}
  \end{center}
\end{figure}
\begin{figure}
  \begin{center}
    \leavevmode
    \includegraphics[width=1.0\columnwidth]{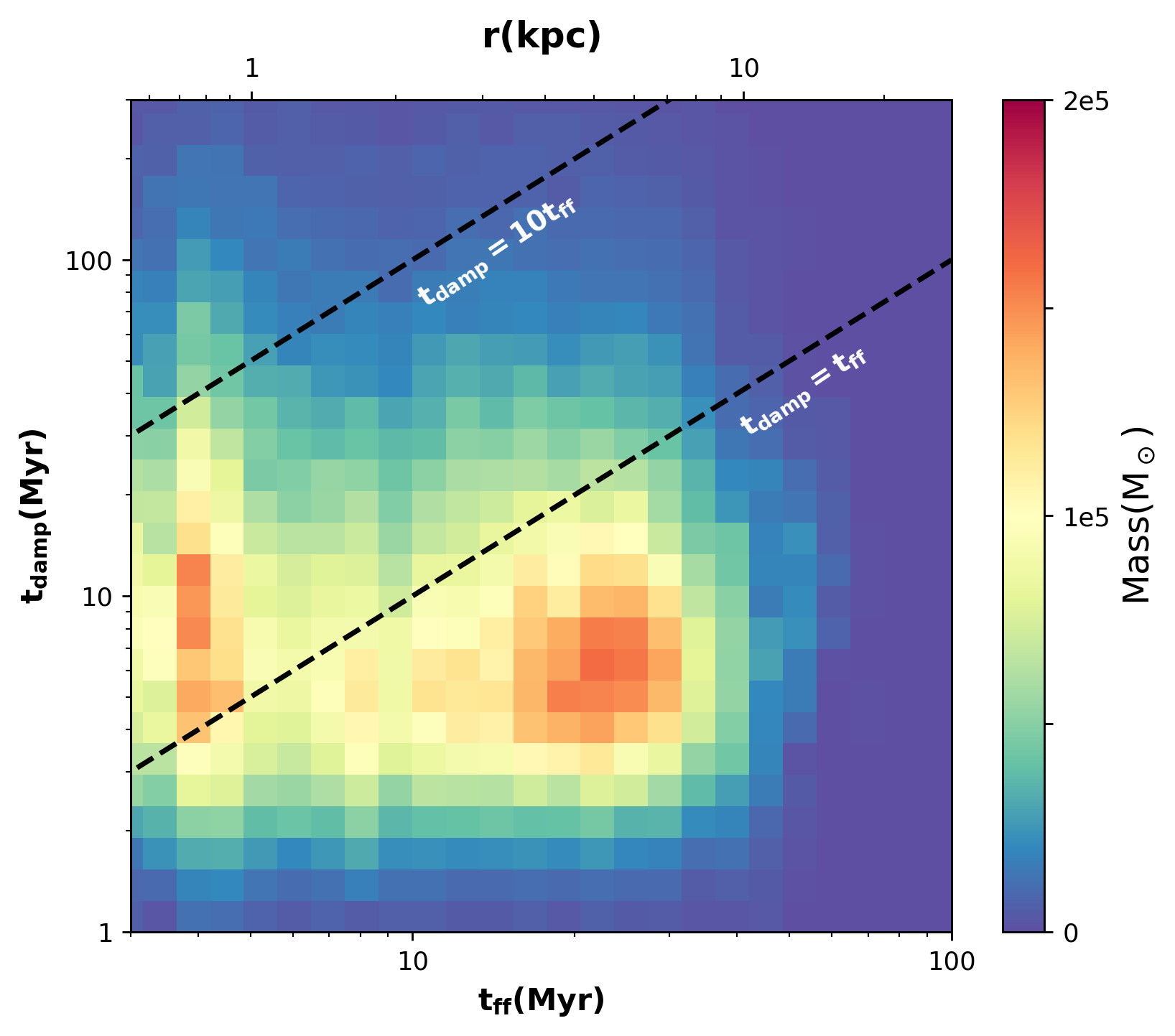}
       \caption[]{The damping time vs. free-fall time ($t_{\rm damp}$ vs. $t_{\rm ff}$) phase plot of the cold gas in \kmag run. Color represents the cold gas mass. The distribution is averaged over snapshots sampled every $2\uu{Myr}$ within the first $2.5\uu{Gyr}$. Dashed lines denote the $t_{\rm damp}=t_{\rm ff} $ and $10t_{\rm ff}$. The upper horizontal axis shows the radii according to the dynamical time.
       }
     \label{fig:magtime}
  \end{center}
\end{figure}
\begin{figure}
  \begin{center}
    \leavevmode
    \includegraphics[width=0.8\columnwidth]{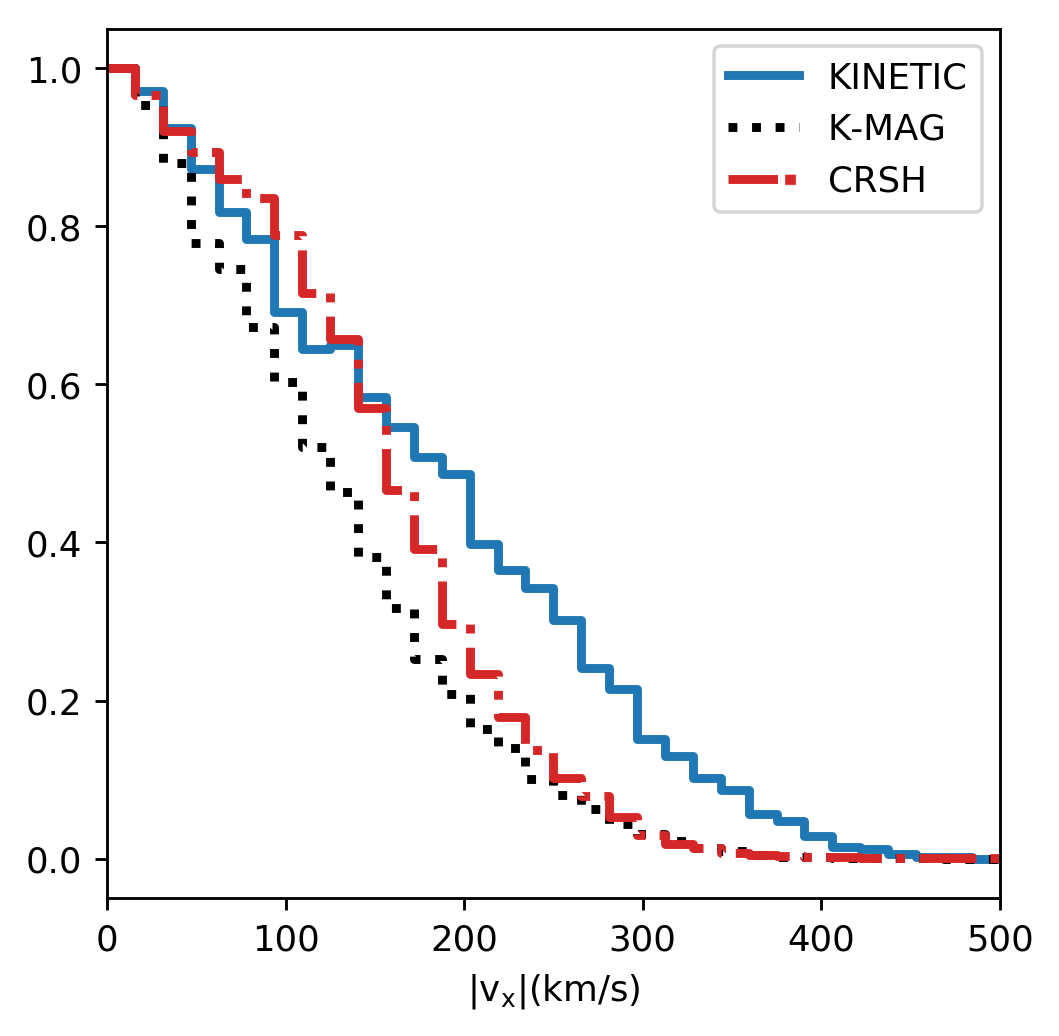}
       \caption[]{Distribution of the magnitude of the cold gas ($T<10^{5}\uu{K}$) velocity along $x-$axis in KINETIC, KMAG, and CRSH run. The three distributions are weighted by gas mass and averaged over the first $1.2\uu{Gyr}$.
       }
     \label{VelocityHistogram}
  \end{center}
\end{figure}
\indent
We can now quantify the magnitude of the tension force in the vicinity of the clouds and estimate the timescale over which it could significantly affect the kinetic energy of the cold gas clumps. The magnetic tension force density is
\begin{equation}
    \mathbf{f_t} =  \frac{1}{4\pi}\nabla\cdot(\mathbf{BB})
\end{equation}
We found that, on average, 75 per cent of the cold gas mas has ${\bf f_{\rm t}}\cdot {\bf u_g}<0$, i.e., that the magnetic tension force does negative work on most of the cold phase. This cold gas can thus be decelerated within a short timescale due to this tension force. [Fig.\ref{fig:tension} shows the time-averaged distribution of the cold gas in the \kmag case on the plane defined by the kinetic energy density $\frac{1}{2}\rho u_g^2$ and the magnetic tension power density $-\mathbf{f_t}\cdot {\bf u_g}$. Color represents the cold gas mass. ] [Consider deleting Fig.\ref{fig:tension} because we don't care about the exact values of kinetic energy / tension power.] We define the ``damping time'' as $t_{\rm damp}=-0.5\rho u_g^{2}/(\mathbf{f_t}\cdot {\bf u_g})$. This is the timescale over which the work done by the magnetic tension force uses up all the kinetic energy of the cold gas. The white dashed lines in Fig.\ref{fig:tension} denote $t_{\rm damp}=1, 10, 100\uu{Myr}$ and $1\uu{Gyr}$, respectively. The damping time of bulk of the gas is about 10 Myr, which is much shorter than the duration of the simulations and shows that magnetic forces have enough time to significantly affect the dynamics of the cold gas and its morphology.
Fig.\ref{fig:magtime} shows the time-averaged distribution of cold gas on the plane defined by $t_{\rm damp}$ and the free-fall time, $t_{\rm ff}=\sqrt{2r/g(r)}$. The dashed lines in Fig.\ref{fig:magtime} denote $t_{\rm damp}=t_{\rm ff}$ and $10t_{\rm ff}$. Most of the cold gas resides in the region where and $t_{\rm damp}<10t_{\rm ff}$, which implies that the magnetic forces can significantly alter the orbits of filaments and lead to dramatic enhancement in the rate of momentum extraction via magnetic breaking, which explains why \kmag run never transitions to the disk phase. We note that the mass concentration at $t_{\rm ff}\approx 4\uu{Myr}$ in Fig.\ref{fig:magtime} is due to the residual angular momentum of the cold gas along the $z-$direction. This cold gas forms transient disk with significantly sub-Keplerian velocity and is quickly accreted.
\\
\indent
Significant magnetic tension forces that are anti-parallel to the velocity vectors of the cold gas not only affect the morphology of that gas phase but also modify its velocity distribution. Fig.\ref{VelocityHistogram} shows the average distribution of the magnitude of the cold gas velocity along the $x-$axis. To make an equivalent comparison between the runs, we only include the data for the first $1.2\uu{Gyr}$, during which both the \kin and \kmag runs experience similar cyclic evolution. This figure clearly demonstrates that the cold gas in \kin run (solid blue line) has a high velocity tail that is absent in the \kmag case (dotted black line).
\subsubsection{Ambipolar diffusion of magnetic fields}
The impact of magnetic tension on the cloud velocity can be reduced when the magnetic fields are allowed to slide past the clouds. This can happen as a result of ambipolar diffusion since the clouds are only partially ionized. In order to assess the impact of this process, we estimate the ambipolar diffusion timescale on which the field sliding occurs, $t_{\rm ad}\sim L^{2}\nu_{\rm in}/u_{\rm A}^{2}$, where $L$ is the filament width, $u_{\rm A}=B/(4\pi\rho_{i})^{1/2}$ is the Alfv{\'e}n speed, $\rho_{i}$ is the ion mass density, and $\nu_{\rm in}\sim 10^{-9}T^{1/2}_{2}n_{n}$ s$^{-1}$ is the ion-neutral collision frequency \citep{dePontieu2001}, where $n_{n}$ is the number density of neutrals and $T_{2}=T/10^{2}$K, where $T$ is the cold phase temperature. We can compare this timescale to the dynamical time $t_{\rm dyn}\sim R/u$, where $R$ is the distance of a cloud from the center and $u$ is its velocity,
\begin{equation}
\frac{t_{\rm ad}}{t_{\rm dyn}}\sim 1.3\times10^{-8}x (1-x)T_{2}^{1/2} \frac{m_{p}n_{H}^{2}L^{2}}{B^{2}}\frac{u}{R} ,    
\end{equation}
where $x$ is the ionization fraction, and $n_{H}$ is the number density of hydrogen. In realistic filaments most of the gas mass is likely to in the molecular form \citep{Ferland2008,Ferland2009}. Assuming that, for low ionization, recombination are balanced by CR ionization, we can estimate the ionization fraction as $x\sim K/n_{n}^{1/2}$, where $K\sim 10^{-5}$cm$^{-3/2}$ \citep{McKee1993,Padoan2000}. This estimate for the ionization fraction assumes ionization rate $\zeta_{H}=10^{-17}$s$^{-1}$ \cite{Spitzer1978}. Since this ionization rate is uncertain and can be larger (e.g., \citet{Indriolo2010,Indriolo2018}), and $K\propto\zeta_{H}^{1/2}$, the above estimate for the ionization fraction is likely a lower limit. For typical, if somewhat conservative, filament parameters 
$R\sim 10$ kpc, $L\sim 10$ pc \citep{Fabian2008}, $u\sim 200$ km $s^{-1}$, $n_{H}\sim 2\times 10^{4}$ cm$^{-3}$, $T\sim 10^{2}$ K, and $B\sim 50\mu$G (see \citet{Werner2013} for the estimates of the last three quantities), we obtain $t_{\rm ad}/t_{\rm dyn}\sim 1.4\times 10^{2}$. This implies that ambipolar diffision is unlikely to lead to significant magnetic line slippage past the cold filaments. While detailed predictions for the filament powering rates are beyond the scope of this work, we also estimated that unlike other mechanisms proposed to significantly contribute the filament powering (e.g., \citet{Fabian2011,Ferland2008,Churazov2013,Ruszkowski2018}), the ambipolar drift heating rate is unlikely to account for the observed amount of power emitted by the H$\alpha$ filaments, which is consistent with relatively long ambipolar diffusion timescales. These conclusions become even stronger for the case of filaments composed predominantly of H$\alpha$-emitting gas of typical density of $\sim$ 30 cm$^{-3}$ and $T\sim 10^{4}$K \citep{Werner2013}.

\begin{figure*}
  \begin{center}
    \leavevmode
    \includegraphics[width=0.95\textwidth]{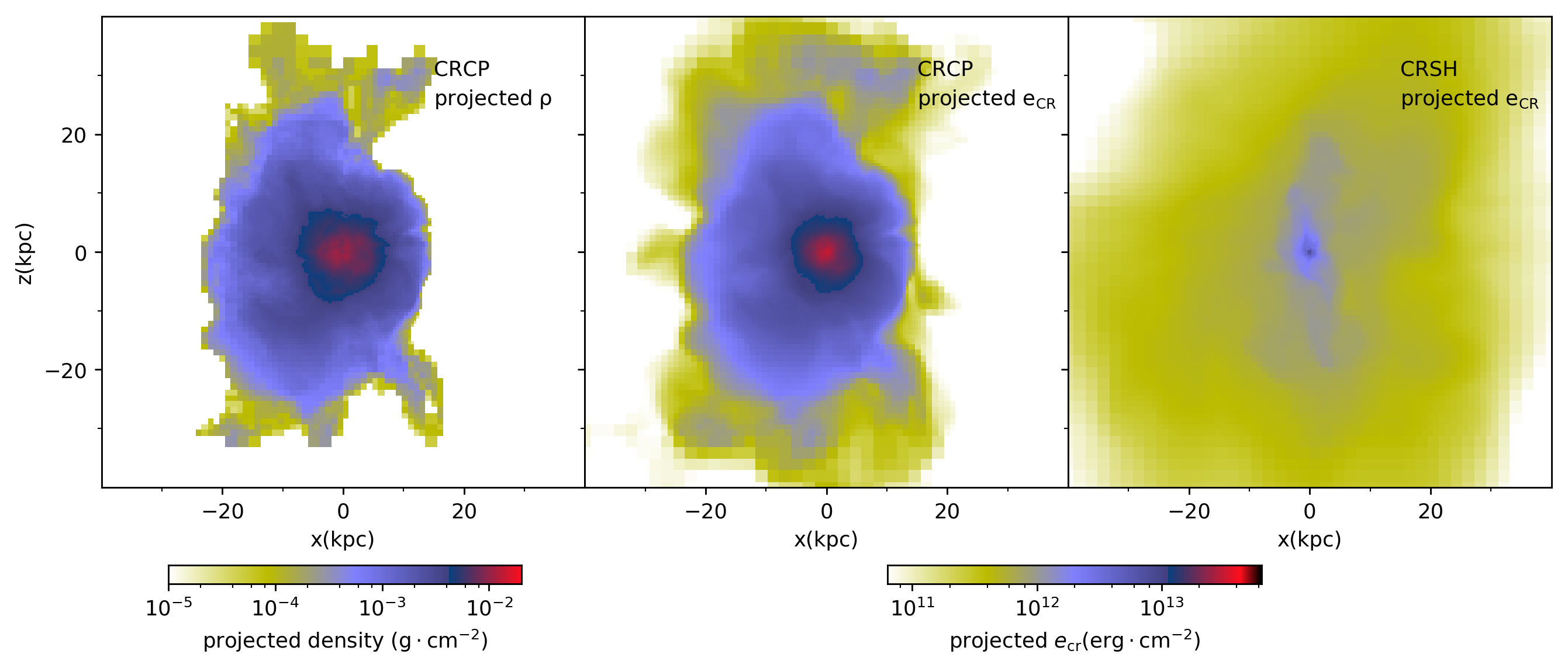}  
       \caption[]{Projected cold gas ($T<10^{5}\uu{K}$) density (left) and CR energy in the \crcp (middle) and \crsh (right) cases at $t=1.43\uu{Gyr}$. The projections are along the $y$-axis and cover the central $80\uu{kpc}$-wide cubic region. }
     \label{fig:8}
  \end{center}
\end{figure*}

\subsection{AGN feedback including magnetic fields and CRs}
As mentioned in the introduction, AGN jets inflate cavities in the CGM that can be dominated by CRs. CR simulations of AGN feedback \citep{2019ApJ...871....6Y} suggest that such CR-filled bubbles could uplift more cold thermally unstable gas because they have lower inertia and flatten in the direction perpendicular to the radial direction, which increases their cross section. These short-duration simulations also show that there is not enough time to transfer the CR energy from the bubbles to the ambient CGM. We now extend the results presented in the above sections to include CR-dominated bubbles inflated by AGN jets in order to study their long-term impact on the CGM. 
\subsubsection{Simulations without CR streaming}
We begin by including CRs and their losses via Coulomb and hadronic interactions with the CGM (run CRCP). These processes heat the thermal CGM. The \crcp run does not exhibit a cycling behavior seen in the \kin and \kmag cases. Instead, it suffers from the cooling catastrophe despite the fact that the AGN continues to inject energy. Cold gas piles up in the center since the early times in the simulation ($t\approx120\uu{Myr}$) and forms an overdense region that gradually grows with time (see third column in Fig.\ref{fig:1}). The cold gas mass increases monotonically as shown in the top panel of Fig.\ref{fig:2}. At $t\approx1.4\uu{Gyr}$, the mass of the cold gas reaches $\sim10^{10}\uu{M_\odot}$. This is two orders of magnitude more than typical values seen in elliptical galaxies. Therefore, we stop the simulation at $t\approx1.4\uu{Gyr}$. One fundamental difference between the \crcp case and the \kin and \kmag cases is that the energy transfer from the bubbles to the CGM proceeds at different rates. While in the \kin and \kmag cases the thermal bubble gas eventually mixes with the surrounding CGM and instantly transfers its energy to the CGM, in the \crcp case the CRs that come into contact with the CGM due to mixing need time to transfer their energy to the CGM via Coulomb and hadronic interactions. Assuming that the bubbles are in pressure equilibrium with the CGM, we can estimate the heating timescale $t_{\rm heat}$ using Eq. 6 and 7. For typical conditions in the cool core at 10 kpc away from the center ($n_{p}\sim 10^{-2}$ cm$^{-3}$, $T\sim 10^{7}$ K), we obtain $t_{\rm heat}\sim 1$ Gyr, which is a significant fraction of the simulation duration and suggests that the energy transfer from CRs to the CGM should be inefficient. \\
\indent
Fig.\ref{fig:8} shows the projected distribution of the cold gas density (left panel) and CR energy density in the \crcp run (middle panel) at $t\approx 1.4\uu{Gyr}$. The two distributions are very similar, which indicates that the cooling gas from the outer parts of the cool core flows in and traps the CR fluid in the region occupied by the cold gas. Although energetic CRs are co-spatial with the cold gas, the inefficient CR heating is unable to reheat the cold gas, which has excessively short cooling times. Consequently, cold gas accumulates in the center and the \crcp run suffers from a cooling catastrophe. Similar result was found in the context of cluster simulations \citep{Ruszkowski2017}. 

\subsubsection{Simulations with CR streaming and heating}
We now extend the CR simulations of AGN feedback discussed above to include the effect of the CR streaming instability (\crsh run). That is, we include CR streaming with respect to the thermal gas and the associated CR heating of the CGM. The right panel in Fig.\ref{fig:8} shows the projected distribution of the CR energy density in the \crsh run. This figure clearly demonstrates that the spatial distribution of CRs can be much broader in this case compared to the one where CR transport is neglected (\crcp run; middle panel). 
\begin{figure*}
  \begin{center}
    \leavevmode
    \includegraphics[width=0.9\textwidth]{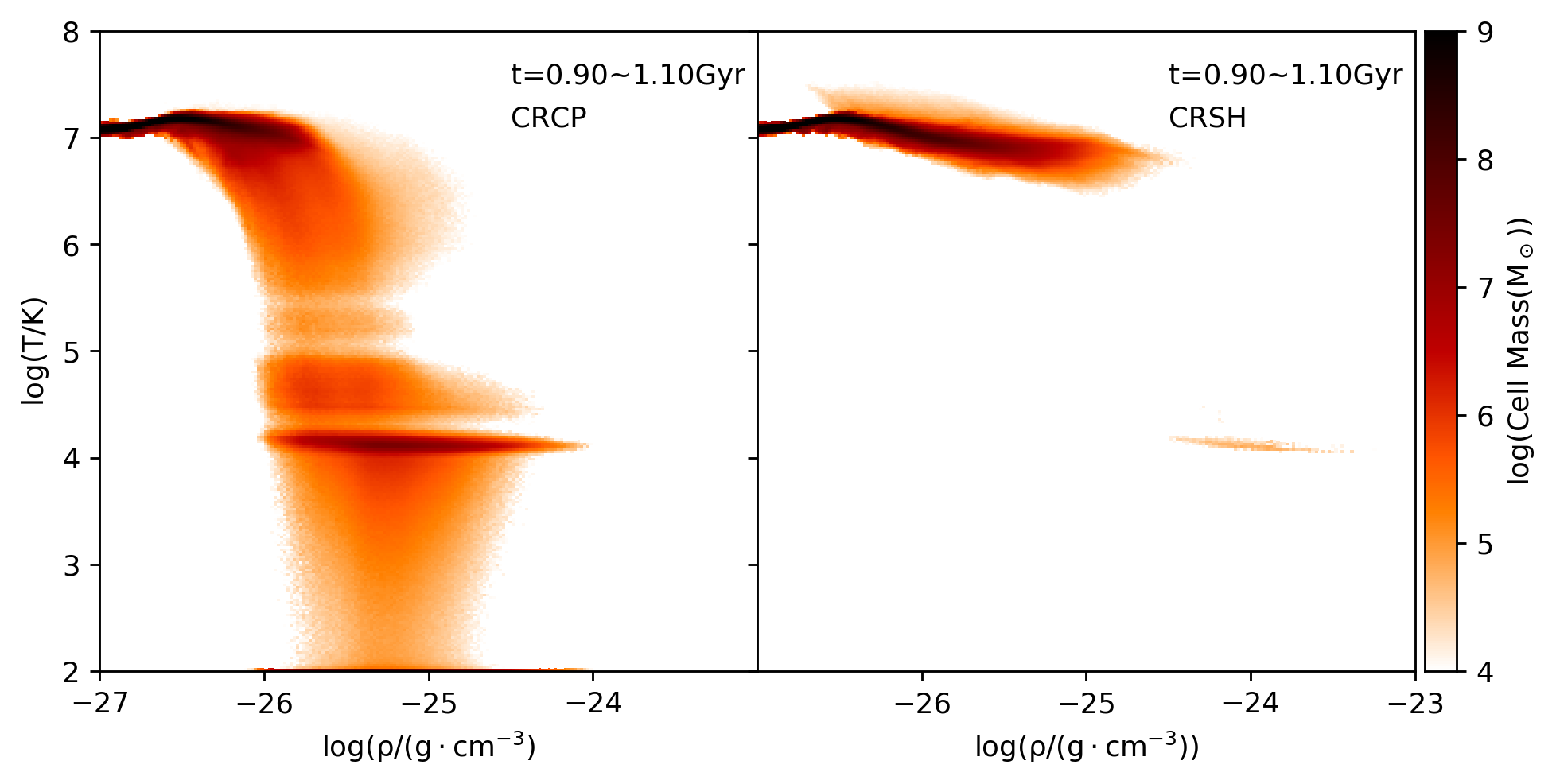}  
       \caption[]{The average temperature - density phase diagrams plotted over $t=0.9\sim1.1\uu{Gyr}$ in CRCP (left) and CRSH run (right). The averaging method is the same as that used by Fig.\ref{fig:3}. Color shows the gas mass.}
     \label{fig:4}
  \end{center}
\end{figure*}
The better dispersal of CRs throughout the CGM in the \crsh case compared to \crcp case leads to more efficient heating as the CRs can come into contact with more hot thermal gas. It is evident from the comparison of the middle and right panels in Fig.\ref{fig:8} that the energy density of CRs is lower in the \crsh case compared to \crcp case. 
The decrease in the CR energy density is due to a combination of adiabatic expansion and streaming of CRs and the additional heating of the gas associated with the CR streaming instability. This heating timescale is $t_{\rm heat, A}\sim 3L/u_{A}$, where $L$ is the scalelength of the CR gradient where significant heating takes place. 
Periods of the most significant CR heating coincide with the AGN outbursts when the pressure gradients of CR escaping from the jet regions are the strongest, and there is relatively little heating between AGN episodes when the CRs have had time to disperse substantially.
Depending on the steepness of the CR pressure gradient, $t_{\rm heat, A}$ can be significantly shorter than the heating timescale $t_{\rm heat}$ corresponding to Coulomb and hadronic losses described above (e.g., $\sim5\times 10^{7}$yr for $L\sim1$ kpc and assuming $\beta = 10^{2}$).
Similar effect was observed in the context of CR simulations of AGN feedback in galaxy clusters \citep{Ruszkowski2017}. Consequently, the cooling catastrophe that was present in the \crcp case does not occur. This is illustrated in Fig.\ref{fig:4} that shows the temperature versus gas density phase space. In the left panel, that corresponds to the \crcp run, large quantities of dense and low temperature gas can be seen, whereas only a relatively very small amounts of precipitating gas can be seen in the right panel that corresponds to the \crsh case. Similarly to the \kmag case, the \crsh case also exhibits cyclic behavior that was absent in the \crcp run (See orange line in Fig.2). Finally, the impact of the magnetic fields on the precipitating clouds in the \crsh case is also reminiscent of that seen in the \kmag run (see right panel in Fig. 4). Consequently, the distribution of cold cloud velocities in the \crsh is also similar to that corresponding to the \kmag case (see red line in Fig. 8) and the cold disk surrounding the SMBH does not form, which is in agreement with observations that suggest that such disks are seen in a relatively small number of systems.\\
\begin{table}
\caption{Components of AGN energy}
\centering
 \begin{tabular}{ccccccccc} 
\hline
name  & $E_{\rm agn}({\rm erg})$  &  $E_{\rm cr}({\rm erg})$
& $E_{\rm cr, used}({\rm erg})$& $\frac{ E_{\rm cr, used}}{f_{\rm cr}E_{\rm AGN}}$ \\
 \hline
 \crcp &  $2.41\times10^{59}$ &  $6.28\times10^{58}$ &
 $1.30\times10^{59}$ & 84\% \\
 
  \crsh &  $1.81\times10^{59}$ &  $2.33\times10^{58}$ &
  $1.22\times10^{59}$ & 67\% \\
 \hline
 \end{tabular}
\end{table}
\indent
That the evolution of the \crcp run differs so substantially from that seen in the \crsh run suggests that CR streaming plays a decisive role in determining whether the CR-dominated AGN jets could efficiently deliver the energy to the CGM. To better understand how efficiently the AGN energy is supplied to the CGM, we analyze the following quantities in these two runs: total energy released by the AGN $E_{\rm agn}$, total energy of the CR fluid that remains in the simulation box $E_{\rm cr}$, and the CR energy transferred to the gas $E_{\rm cr, used}=f_{\rm cr}E_{\rm AGN}-E_{\rm cr}$ computed assuming negligible CR outflow through the box boundary. 
In principle, $E_{\rm cr,used}$ should exceed the energy that is actually used to heat the gas because a fraction of CR energy that lost due to hadronic collisions is removed from the system via gamma ray and neutrino emission. However, we verified that the hadronic energy losses in both cases are small compared to other terms and therefore approximately all of $E_{\rm cr,used}$ is utilized to heat the gas.
Comparison of these quantities for the \crcp and \crsh runs at $t\approx1.2\uu{Gyr}$ (see Table 2) reveals that: (i) $E_{\rm agn}$ is larger in \crcp run than that in \crsh run. This is likely caused by the weak coupling of the AGN energy to the gas. The AGN is trying to compensate for this weak heating efficiency by injecting more energy into the CGM. One manifestation of this is the fact that while in the \crcp run the AGN is continuously injecting energy, the \crsh run undergoes a quiescent phase at $t\approx750\uu{Myr}$ during which AGN releases zero energy as the atmosphere is relatively close to being in the state of global thermal equilibrium (the solid orange line in the middle panel of Fig.2); (ii) the above finding is consistent with the observation that, due to the additional streaming heating, CR energy is more efficiently utilized in \crsh run, i.e., $E_{\rm cr, used}/(f_{\rm cr}E_{\rm AGN}) \approx84\%$ and $\approx67\%$ in the \crsh and \crcp cases, respectively.\\
\indent
The total CR energy transferred to gas is slightly higher in the \crcp (as the AGN is trying to compensate for the low heating efficiency), but the heating in this case still fails to balance cooling globally. However, the ability of the AGN to shut off catastrophic cooling of the atmosphere depends not only on the total amount of energy transferred to it but also on how uniform the heating is distributed and how much of the cooling gas can come into contact with CRs. As mentioned above, the spatial distribution of CRs is broader in the \crsh case which, together with the fact that the amount of the injected energy is better utilized in this case, explains why the cooling catastrophe is prevented when CR streaming effects are included. \\
\indent
The better utilization of the AGN energy is also evident from the bottom panel in Fig.\ref{fig:2}. This figure shows the evolution of the total AGN energy $E_{\rm AGN}$. The smallest values of $E_{\rm AGN}$ correspond to the \crsh case, i.e., the smallest values of AGN energy suffice to offset cooling in a time average sense and prevent cooling catastrophe. 

\section{Conclusions}
We performed MHD simulations of AGN feedback in elliptical galaxies including the effects of CR heating and streaming. In these simulations we followed the evolution for a long period of time (Gyrs). We found that non-ideal hydrodynamical effects have a profound impact on the evolution of elliptical galaxies. In other words, the evolution of the systems on 100 kpc scales depends sensitively on the physics governing the ``microscopic'' scales relevant to plasma physics. Specifically, we showed that
\begin{enumerate}
    \item 
    in the purely hydrodynamical case, the AGN jet initially maintains the atmosphere in global thermal balance. However, local thermal instability eventually leads to the formation of massive cold disk in the vicinity of the central SMBH. This is a generic finding. Similar conclusion was found in other grid-based and smooth particle hydrodynamics simulations performed by other groups. Once the disk has formed, it continues to feed the central SMBH and the jet begins to overheat the central regions of the atmosphere and does not prevent fast cooling of the gas in the direction perpendicular to it close to the center. Thus the AGN-to-CGM energy coupling is weak due to the inflow-outflow geometry very close to the SMBH. This leads to further disk growth and excessive buildup of the cold gas in the very center, which is in disagreement with the observations; 
    
    \item including high-$\beta$ magnetic fields prevents the formation of the cold disks. The removal of the cold gas from the central regions is due to local B-field amplification in the cold precipitating gas and increased magnetic breaking acting on the cold blobs. The local field amplification occurs via flux freezing during the development of local thermal instability, which is followed by stretching of the fields caused by the motion of the cold gas clumps that dynamically decouple from the hot gas. The B-field line stretching leads to filamentary/cometary appearance of the cold clouds that do not fall ballistically onto the center. In a statistical sense, the magnetic tension vectors are anti-parallel to the cold cloud velocities, which leads to magnetic breaking of the cold clouds. Even though the magnetic field in both the cold and hot phases remains dynamically weak throughout the evolution, the tension forces acting on the cold clouds are capable of affecting their dynamics on timescales much shorter than the evolutionary timescale of the system. We also find that magnetic fields can narrow the velocity distribution of the cold phase (compared to that observed in the purely hydrodynamical case);
    
    \item when plasma composition in the AGN jets is dominated by CRs, and CR transport is neglected, the atmospheres exhibit cooling catastrophes due to the inefficient heat transfer from the AGN to CGM despite Coulomb/hadronic CR losses being present. In this case, CRs continue to accumulate in the center due to the fact that CRs do not come into direct contact with sufficiently large fraction of the radiatively cooling atmosphere. CR energy is not utilized very efficiently due to the lack of transport and the fact that typical CR cooling timescales associated with Coulomb and hardonic interactions are a substantial fraction of the system lifetime;
    
    \item including CR streaming and heating restores the agreement with the observations, i.e., cooling catastrophes are prevented, the gas does not overheat, and massive cold central disks do not form. The AGN power is reduced as its energy is utilized efficiently. 
    
\end{enumerate}

\section*{Acknowledgements}
MR acknowledges NSF grant AST 1715140. HYKY acknowledges support from NASA ATP (NNX17AK70G) and NSF (AST 1713722). We thank Yuan Li for useful discussions. MR thanks Eric Blackman, Jim Drake, Peng Oh, Christoph Pfrommer, and Ellen Zweibel for very enlightening discussions at KITP. This work used the Extreme Science and Engineering Discovery Environment (XSEDE, \citealt{xsede}), which is supported by National Science Foundation grant number ACI-1548562. We acknowledge the XSEDE computational grant TG-AST180063. MR thanks the Aspen Center for Physics and the Kavli Institute for Theoretical Physics at the University of California at Santa Barbara, where part of this work was performed. The Aspen Center for Physics is supported by National Science Foundation grant PHY-1607611. This research was supported in part by the National Science Foundation under Grant No. NSF PHY-1748958.

\bibliographystyle{mnras}
\bibliography{template}

\bsp	
\label{lastpage}
\end{document}